\documentclass[%
 reprint,
 prc,
 amsmath,amssymb,
 aps,
 lengthcheck,
floatfix,
]{revtex4}

\usepackage{graphicx}
\usepackage{dcolumn}
\usepackage{bm}
\usepackage{hyperref}
\usepackage[utf8]{inputenc}
\usepackage{multirow}
\usepackage{soul}
\usepackage{float}
\usepackage{graphicx}
\usepackage{times}
\usepackage[normalem]{ulem}
\usepackage{color}
\usepackage{cellspace}
\usepackage{ulem}

\newcommand{\be}{\begin{equation}}
\newcommand{\ee}{\end{equation}}
\newcommand{\bea}{\begin{eqnarray}}
\newcommand{\eea}{\end{eqnarray}}

\newcommand{\comment}[1]{}
\renewcommand\sout{\bgroup \color{red} \ULdepth=-.5ex \ULset}
\def\simge{\mathrel{\rlap{\raise 0.511ex
     \hbox{$>$}}{\lower 0.511ex \hbox{$\sim$}}}}
\def\simle{\mathrel{\rlap{\raise 0.511ex
      \hbox{$<$}}{\lower 0.511ex \hbox{$\sim$}}}}

\begin{document}

\title{Can we decipher the composition of the core of a neutron star?}

\author{C. Mondal$^1$}
\email{mondal@lpccaen.in2p3.fr}
\author{F. Gulminelli$^1$}

\affiliation{$^1$Laboratoire de Physique Corpusculaire, CNRS, ENSICAEN, UMR6534, Université de Caen Normandie,
F-14000, Caen Cedex, France}

\date{\today}

\begin{abstract}
General relativity guarantees a unique one-to-one correspondence between static
observables of neutron stars (NSs) accessible by multi-messenger astronomy, 
such as mass-radius or tidal deformability, and the equation 
of state (EoS) of beta equilibrated matter. It is well known that these static 
properties are not enough to discern conventional NSs from hybrid stars.  
However, if one assumes that hadrons present in the neutron star core are only 
neutrons and protons, the lepton fraction is in principle determined 
unequivocally by the condition of chemical equilibrium. Using a simple analytical 
method based on a polynomial expansion of the EoS, we show that multiple 
solutions are possible to the beta-equilibrium equation, leading to a characteristic 
indetermination on the composition of the interiors of NSs, even in the purely 
nucleonic hypothesis. We further show that additional empirical information on 
symmetric matter at high densities are not very efficient to pin down the composition, 
if uncertainties on measurements are accounted for. We conclude that constraints 
on the symmetry energy at high densities only, can make meaningful impact to decipher
the composition of neutron star core. Our results give a lower limit to the 
uncertainty on the NS core composition that can be obtained with astrophysical 
and terrestrial experiments.

\end{abstract}

\maketitle



\section{Introduction}

Neutron stars are among the most dense systems existing in the universe. Understanding
the composition of their core will help us to peek at the behavior of matter at extreme
density conditions. Unprecedented progress was achieved in multi-messenger astronomy in the
last decade through quantitative measurements of neutron stars (NS) properties, such as
mass measurements by radio astronomy through Shapiro delay
\cite{Demorest10, Antoniadis13}, joint mass-radius determination by
NICER using X-ray timing data \cite{Riley19, Miller19, Riley21, Miller21} or the tidal
polarizability extracted from the gravitational wave signal by LIGO-Virgo
collaboration \cite{Abbott17, Abbott18, Abbott19, LIGO15, Acernese14}.

A straightforward link between the observations and the underlying microphysics
can be established due to a one-to-one correspondence between the
static properties of NS and the EoS of matter under the realm of general relativity
\cite{Hartle67}. The behavior of the dense matter EoS can therefore be extracted with
controlled uncertainty from the astrophysical observations within minimal assumptions
using Bayesian techniques  \cite{Steiner13, Margueron18a, Margueron18b, Zhang18, Lim19,
Tsang20, Traversi20, Biswas21, Essick21}.  However, a major persisting challenge consists
in connecting this empirically determined EoS with the internal properties of dense matter,
notably to bring to light the existence (or absence) of hyperonic degrees of freedom
and the deconfined quark matter in the core of neutron stars \cite{Burgio21}. 
Because of the well known ``masquerade'' phenomenon \cite{Alford05}, hybrid stars including 
a quark core can exhibit a mass-radius [$M$-$R$]  relationship very similar to the one 
obtained for a star made of purely nucleonic matter, see \cite{Baym18} for a recent review.

The discrimination between confined and deconfined matter in the NS core is clearly 
of foremost importance for our understanding of the QCD phase diagram. However, 
even in the simplified assumption of a purely nucleonic composition, a quantitative 
knowledge on the composition is of utmost importance. Indeed,
the electron fraction in the star core is a crucial input both for differentiating the different
nuclear theories, and to correctly model  dynamical processes such as pulsar glitches, 
cooling and mergers \cite{Oertel17,Potekhin18,Montoli20}. { Extraction of information 
on the composition from these dynamical processes is in principle possible, but it is clearly 
limited by the many microscopic and macroscopic unknowns in the complex theoretical modelling.  }

In the conventional density functional approaches, the energy
functional fixes the composition \textit{a priori}, when one solves the
$\beta$-equilibrium condition equation to obtain the EoS of NS matter.
Since the latter has a one-to-one connection to the NS  $M$-$R$ relation,
it is systematically assumed that, in the absence of exotic degrees of freedom, 
the uncertainty in the composition only arises from the error bar on the $M$-$R$ relation.   
However, we will demonstrate in this paper that 
the correspondence between the
NS composition and the EoS, as it can be accessed through astrophysical
observations, is not unique, even in the purely nucleonic hypothesis. This
surprising result is due to the
existence of multiple solutions to the $\beta$-equilibrium
equation, which is analytically proved  using two different realizations of the same EoS. 
The result is reinforced by a Bayesian analysis hypothesizing controlled uncertainties 
on the pressure of $\beta$-equilibrated matter at distinct number densities. 
Further, we made the analysis more realistic by extracting the EoS from an astrophysical
measurement through the inversion of the Tolman-Oppenheimer-Volkoff (TOV)
equation of hydrostatic equilibrium. We
show that the propagation of uncertainties is such that the composition above twice
the saturation density to be fully unconstrained, even if the EoS was pinned down 
very precisely. We tested independent complementary information from laboratory
experiments at suprasaturation densities to extract meaningfully the
proton fraction (or equivalently electron fraction) in the NS core and, consequently, 
the nuclear matter energy functional.

The plan of the paper is as follows.  In Section \ref{sec:formalism}, we give the 
theoretical formalism based on the inversion of the $\beta$-equilibrium equation 
through an analytical polynomial expansion of the EoS in  Section \ref{sec:inverting}, 
and different Bayesian schemes in Section  \ref{sec:Bayes}. Section \ref{sec:results}  
comprises the results from two approaches, first on the inversion of the baryonic 
EoS in Section \ref{sec:prho}, and on a more realistic case of the $M$-$R$ relation 
in Section \ref{sec:MR}. We explore the potential impact of laboratory constraints 
of the high density regime at fixed proton fraction, on the $\beta$-equilibrium 
composition in Section \ref{sec:lab}. A brief summary and conclusions are drawn 
in Section \ref{sec:concl}.

\section{Inverting the $\beta$-equilibrium equation}\label{sec:formalism}
\subsection{Analytical approach}\label{sec:inverting}

We consider purely nucleonic matter, characterized by its baryonic density $n$ and
asymmetry $\delta=(n_n-n_p)/n$, where $n_{n(p)}$ is the neutron (proton) density.
In order to study the possibility of extracting the
composition from the knowledge on the EoS, we take the beta-equilibrium
energy functional from an arbitrary reference nuclear model $e_{th,\beta}(n)$, and
solve the equilibrium equation for the composition $\delta_\beta(n)$:
\be
\mu_n(n,\delta_\beta)-\mu_p(n,\delta_\beta)=\mu_e(n,\delta_\beta). \label{beta}
\ee
Here, the electron chemical potential is simply given by the Fermi energy:
$\mu_e=\sqrt{(3\pi^2n_e)^{2/3}+m_e^2}$ with $n_e=(n_p-n_\mu)$ and $n_\mu$ the net muon
density in the global equilibrium ($\mu_\mu=\mu_e$). The neutron and proton
chemical potentials are given by:
\be
\mu_{n,p}=\frac{\partial (n e_{th})}{\partial n_{n,p}}
=e_{th}+m_{n,p}+n\left.\frac{\partial e_{th}}
{\partial n}\right|_\delta \pm \left.\frac{\partial e_{th}}{\partial \delta}\right|_n, \label{munp}
\ee
where $m_{n,p}$'s are the free nucleon masses.
The quantity $e_{th}(n,\delta)$
is the energy per baryon that we want to
reconstruct from the knowledge on  $e_{th,\beta}(n)$ in a finite number
of density points $n_i, i=1,\dots,N$. Replacing Eq. (\ref{munp}) in Eq. (\ref{beta}) we immediately get:
 \be
 2\left.\frac{\partial e_{th}(n,\delta_\beta)}{\partial \delta}\right|_n=
 \mu_e(n,\delta_\beta)-\Delta m_{np}, \label{eq:beta}
 \ee
with $\Delta m_{np}=m_n-m_p$. To evaluate the l.h.s. of Eq. (\ref{eq:beta}) in order to
extract $\delta_\beta(n)$, we need to parametrize the functional out of the
$\beta$-equilibrium trajectory. A convenient representation that can precisely
reproduce any generic realistic nucleonic functional \cite{Margueron18a},
is given by the so called meta-model approach as,
\begin{eqnarray}
e_{meta}(n,\delta)=t_{FG}^*(n,\delta) +U_0(n)+U_{sym}(n)\delta^2 .
\label{meta}
\end{eqnarray}
Here, $t_{FG}^*$ is a Fermi-gas type kinetic energy term which also takes into account
the density dependence of the effective nucleon masses, and the deviation of
the parabolic approximation of the symmetry energy \cite{Margueron18a};
$U_0(n)$ and $U_{sym}(n)$ can be attributed to the symmetric and asymmetric parts of the
nuclear potential written as a Taylor's expansion in density that, keeping up to 4th order, can be written as,
\begin{eqnarray}
U_{0,sym}(n)=\sum_{k=0}^4 \frac{(v_k)_{0,sym}}{k!}x^k,\ {with,}\ x=\frac{n-n_{sat}}{3n_{sat}},
\end{eqnarray}
where $(v_k)_{0,sym}$ are functions of nuclear matter properties (NMPs)
at saturation density $n_{sat}$ \cite{Margueron18a}. The NMPs needed for the
meta-functional in Eq. (\ref{meta}) are the energy per particle $E_{sat}$,
incompressibility $K_{sat}$, skewness $Q_{sat}$ and stiffness $Z_{sat}$ of
symmetric nuclear matter (SNM), $e_0(n)=e_{th}(n,\delta=0)$; and symmetry energy $E_{sym}$, symmetry slope
$L_{sym}$, symmetry incompressibility $K_{sym}$, symmetry skewness $Q_{sym}$ and
symmetry stiffness $Z_{sym}$ corresponding to the density dependent symmetry energy, 
$e_{sym}(n)=(1/2)\partial^2 e_{th}/\partial \delta^2(n,\delta=0)$,
all evaluated at the equilibrium density of nuclear matter $n_{sat}$.
The NMPs are connected to the successive derivatives of the energy functional 
at saturation in the isoscalar (symmetric matter $e_0$) and isovector (symmetry 
energy $e_{sym}$) sector. For the exact definitions of $(v_k)_{0,sym}$ c.f. 
Eq. (21-31) of Ref. \cite{Margueron18a}.

Over the last three or four decades an enormous amount of theoretical and experimental
works have been devoted to obtain the different nuclear matter properties. For example,
the binding energy data of nuclei very well constrain $E_{sat}$ and $E_{sym}$
\cite{Moller12, Jiang12}, giant monopole resonance energies limit the value of $K_{sat}$
\cite{Khan12, De15} and with some reasonable ambiguities $L_{sym}$ and
$K_{sym}$ can be constrained by many different approaches \cite{Centelles09, Roca-Maza11,
Abrahamya12, Mondal15, Mondal16, Chen14, Chen15, Mondal17, Mondal18, Reed21}.  Since the
lower order NMPs are relatively well constrained, for this analytical exercise we consider
that they are known exactly in Eq. (\ref{meta}), and  take the corresponding values of the
reference model which we are trying to reproduce. However, the rest of the
parameters $Q_{sat,sym}$ and $Z_{sat,sym}$ are not experimentally accessible, and scattered
over a huge range of values across many different nuclear models \cite{Dutra12, Dutra14}.

These higher order parameters can be determined by equating the input $e_{th,\beta}(n)$
to its meta-model representation $e_{meta}(n,\delta_\beta)$, at the asymmetry
$\delta_\beta(n)$ corresponding to the solution of Eq. (\ref{eq:beta}). Using the {\it ansatz}
of Eq. (\ref{meta}) we arrive at the equation,
\begin{eqnarray}
\frac{1}{6}x^3Q_{sat}+\frac{1}{24}x^4Z_{sat}+\frac{1}{6}x^3\delta_\beta^2Q_{sym}+
        \frac{1}{24}x^4{\delta_\beta}^2Z_{sym}\nonumber\\
        =e_{th,\beta}(n)-e_{meta}^0(n,{\delta}_\beta),
\label{matrix_inv}
\end{eqnarray}
where $e_{meta}^0$ is the functional $e_{meta}$ from Eq. (\ref{meta}) obtained by setting the
lower order NMPs from the input functional, and  $Q_{sat,sym}=Z_{sat,sym}=0$.
The simultaneous solution of Eq. (\ref{matrix_inv}) and (\ref{eq:beta}) can be obtained with
a simple iterative matrix inversion by inputting the values of $e_{th,\beta}(n_i)$
corresponding to four distinct density points $i=1,\dots,4$ as:
\begin{equation}
   \left [
   \begin{array}{c}
   Q_{sat}\\ Z_{sat}\\   Q_{sym}\\ Z_{sym}
   \end{array}
\right ]=24
\left [
            \begin{array}{cccc}
4x_1^3 &  x_1^4  &  4x_1^3\delta_1^2 &  x_1^4\delta_1^2\\
4x_2^3 &  x_2^4  &  4x_2^3\delta_2^2 &  x_2^4\delta_2^2\\
4x_3^3 &  x_3^4  &  4x_3^3\delta_3^2 &  x_3^4\delta_3^2\\
4x_4^3 &  x_4^4  &  4x_4^3\delta_4^2 &  x_4^4\delta_4^2
\end{array}  \right ]^{-1}
   \left [
   \begin{array}{c}
   e_1\\  e_2 \\    e_3\\   e_4
   \end{array}
\right ] ,
\label{matrix}
\end{equation}
with $e_i=e_{th,\beta}(n_i)-e_{meta}^0(n_i,{\delta}_i)$, and $\delta_i$ the solution of
the meta-model equivalent of Eq.(\ref{eq:beta}):
 \be
 2\left.\frac{\partial e_{meta}(n_i,\delta)}{\partial \delta}\right|_{n_i}=
 \mu_e(n_i,\delta)-\Delta m_{np}. \label{eq:betameta}
 \ee

The solution of the coupled equations (\ref{eq:betameta}), (\ref{meta}) and (\ref{matrix}) 
fully determines the functional $e_{meta}$ that by construction coincides with the 
reference model in $\beta$-equilibrium, at the four chosen density points. Possible multiple 
solutions of the $\beta$-equilibrium equation can then be sought out, by slightly varying the chosen 
density points, or modifying the initialization of the iterative procedure, 
resulting in different $Q_{sat,sym}$ and $Z_{sat,sym}$. Such multiple solutions 
arise from the indetermination of the high order NMPs $Q_{sat,sym}$ and $Z_{sat,sym}$. If we 
consider that parameters such as $L_{sym}$ and $K_{sym}$ are not sufficiently pinned down as well,  
the procedure of Section \ref{sec:inverting} could also be extended to a higher set of NMPs by 
increasing the number of density points in Eq.(\ref{matrix}). For this reason, in the Bayesian 
approach presented in the next Section, we have also added the relatively poorly constrained 
$K_{sym}$ parameter to our prior. However, we have checked that adding an extra dimension 
to Eq.(\ref{matrix}) including   $K_{sym}$ does not change the results of Section \ref{sec:results}. 

\subsection{Bayesian approach}\label{sec:Bayes}

In the astrophysical context, the EoS of dense matter is not directly measured
at different density points; they are rather inferred from the mass, tidal
polarizability and radius through the structural equations of the star in general relativity, which
imply an integration of the $\beta$-equilibrated EoS over the whole star density profile. Moreover, the
observational quantities are known with finite precision, inducing an
uncertainty that propagates to the EoS, and might
blur the correlation between the EoS and the energy functional and
composition as given by the matrix inversion in Eq. (\ref{matrix}). To account for this limitation,
we attempt a reconstruction of the reference model from the general meta-functional
Eq. (\ref{meta}) through a Bayesian approach, where we sample the prior keeping
fixed the lower order parameters (which are considered as ``known'' from independent
measurements) $E_{sat}, K_{sat}, E_{sym}$ and $L_{sym}$ to the corresponding model
values. {The rest of the parameters $K_{sym}$, $Q_{sat,sym}$ and $Z_{sat,sym}$ are
populated in a Monte-Carlo sampling {with a flat distribution} { as 
$K_{sym}=[-400,200]$, $Q_{sat}=[-1000,1000]$, $Q_{sym}=[-2000,2000]$, 
$Z_{sat}=[-5000,5000]$, $Z_{sym}=[-5000,5000]$, all in units of MeV.}}
These ranges are chosen such as to include
a large number of popular relativistic as well as non-relativistic functionals 
\cite{Margueron18a, Dinh-Thi21}. We have checked that the
arbitrary choice of the reference model does not play any significant role  in the 
qualitative results presented in this paper.

The posterior distributions of the set ${\bf X}\equiv\{K_{sym}, Q_{sat}, Q_{sym},
Z_{sat}, Z_{sym}\}$ of EoS parameters  are conditioned by
likelihood models of the different observations and constraints $\mathbf c $ according
to the standard definition:
\begin{equation}
        P({\mathbf X}|{\mathbf c})=\mathcal N   P({\mathbf X}) \prod_k P(c_k|{\mathbf X}),
\end{equation}
where $ P({\mathbf X})$ is the prior, and $\mathcal N$ is a normalization factor.
Posterior distributions of different observables $Y$ are  calculated by marginalizing over
the EoS parameters  as:
\begin{equation}
P(Y|{\mathbf c})=\prod_{k=1}^M\int_{X_k^{min}}^{X_k^{max}}dX_k \,
        P({\mathbf X}|{\mathbf c}) \delta\left (Y-Y({\mathbf X})\right ),
\end{equation}
where $M=5$ is the number of parameters in the meta-model which are varied.
$Y({\mathbf X})$ is the value of
any observable $Y$ obtained with the ${\mathbf X}$ parameter set along with the fixed lower
order parameters, with $X_k^{\rm min(max)}$ being the minimum (maximum) value in the prior
distribution taken for the analysis.
 The constraints $c_k$ are pseudo-observations from the reference model FSU2. 
We will successively consider different choices for the pseudo-observations, leading to four different posterior distributions:
\begin{itemize}
\item {\it post-1 -} The total equilibrium pressure $P_{th}$ of the reference model is imposed at four different density points, as in the analytical inversion presented in Section \ref{sec:inverting}. No uncertainty is considered on the baryonic density, and the likelihood model is a Gaussian distribution with $\sigma_P/P_{th}=0.1$ .
\item {\it post-2 -} Three hypothetical radii measurements at fixed NS
mass are imposed, where the 1-$\sigma$ uncertainty is arbitrarily fixed to 5\% of  the model values: 
		$R_{1.4M_{\odot}}$, $R_{1.7M_{\odot}}$  and  $R_{2.01M_{\odot}}$.
 \item {\it post-3 -} The pseudo-observations of
radii of post-2 are complemented with a hypothetical SNM pressure observation at $4n_{sat}$ calculated from the reference model with 10\% precision.  
 \item {\it post-4 -} An extra constraint is further imposed on top of post-3 
 as the theoretical value of the symmetry energy at $4n_{sat}$, again with 10\% precision.
 \end{itemize}
 
\section{Results}\label{sec:results}

The determination of the energy functional and matter composition from the inversion of the 
$\beta$ equilibrium information as explained in Section \ref{sec:formalism}
 is illustrated using the relativistic mean field (RMF) model
FSU2 \cite{Chen14} as the reference model, including nucleons, electrons and muons. 
We have repeated all the calculations shown in this paper using the BSK24\cite{Goriely15}, 
the Sly4\cite{Chabanat98}, and the SINPA\cite{Mondal16} models.
Though the quantitative predictions are obviously different, all the qualitative results presented in the paper are independent of the choice of the reference model.
 
{ \subsection{Constraining the pressure of $\beta$-equilibrated matter}\label{sec:prho}}
{To illustrate the existence of multiple solutions to the $\beta$ equilibrium equation, 
and validate the Bayesian approach, we first consider the academic situation of post-1,
where the pseudo-observations  are the values of the FSU2 equilibrium pressure at different 
density points, chosen to be similar $n_i$, $i=1,\dots,4$ used for the analytical inversion in 
Eq.(\ref{matrix}), with an arbitrarily chosen 10\% precision.}

\begin{figure}[h]{}
\includegraphics[height=2.in,width=3.in]{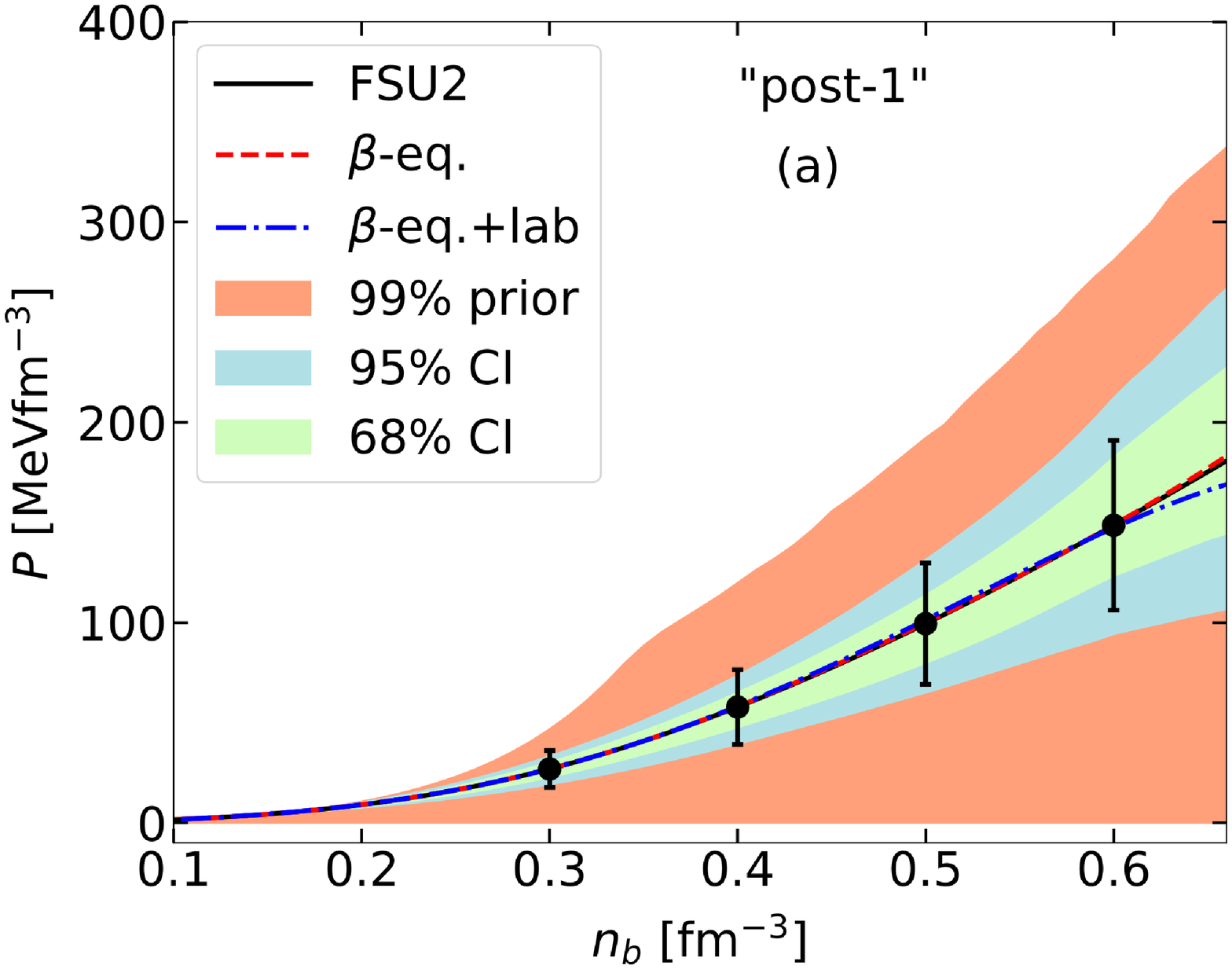}
\includegraphics[height=2.in,width=3.in]{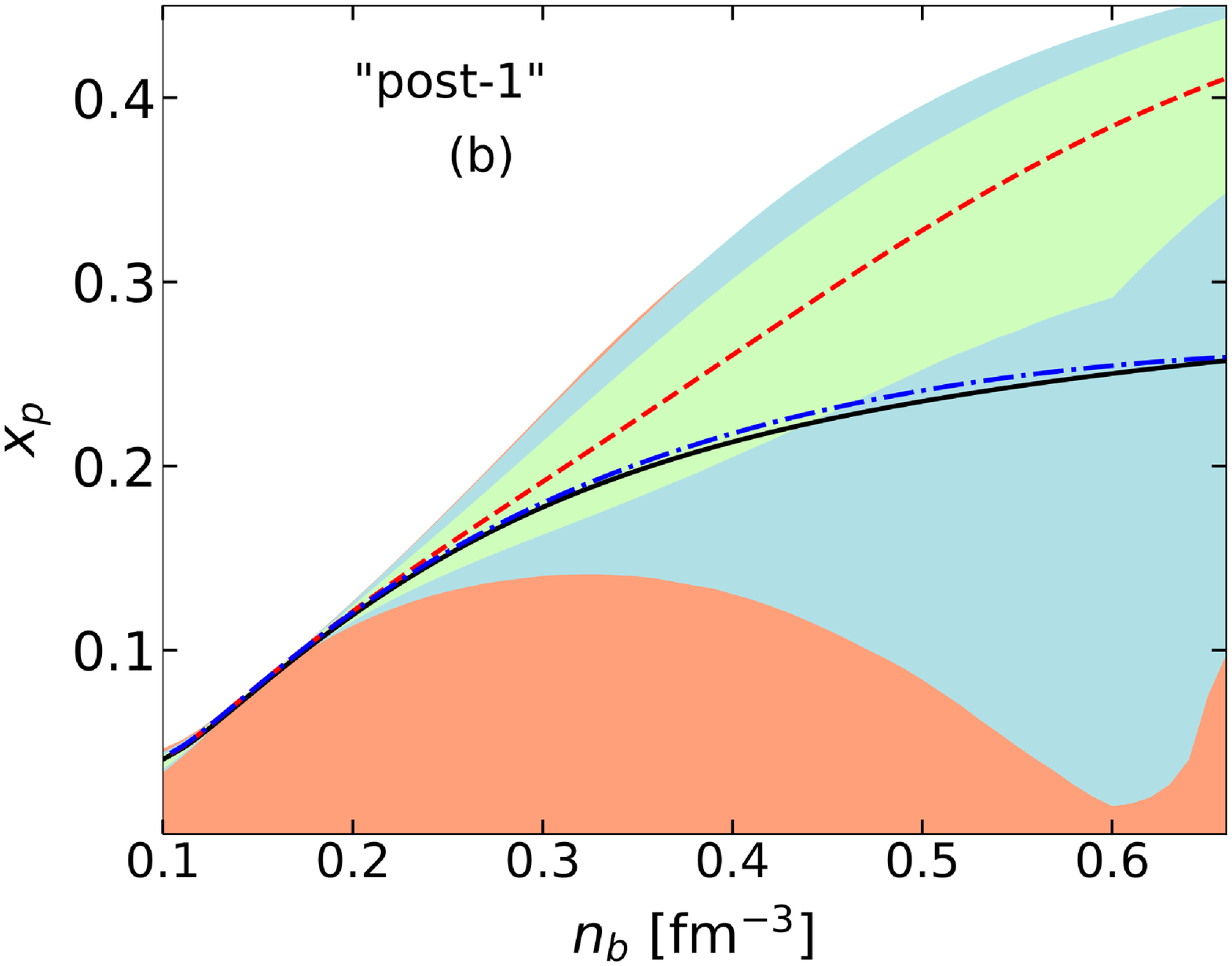}
\caption{\label{post1}
(Color online) Pressure (panel a) and electron fraction (panel b) for $\beta$-equilibrated 
	matter as a function of density for FSU2  and its meta-model equivalents ``$\beta$-eq." 
	and ``$\beta$-eq.+lab" along with its prior and posterior distributions
        with different confidence intervals (CI) calculated from ``post-1"  (see text for details).
}
\end{figure} 
{In Fig. \ref{post1} the NS EoS calculated from FSU2 is given by the black line. The
corresponding red-dashed, and blue dashed-dotted lines give two different meta-model equivalents
of FSU2 produced with Eq. (\ref{matrix}). The red dashed lines (``$\beta$-eq.") are obtained by  
considering initial guess values for the parameters $Q_{sat}=Z_{sat}=Q_{sym}=Z_{sym}=0$,
while the functional described by the blue dashed-dotted line (``$\beta$-eq.+lab") is obtained 
if $Q_{sat}$ and $Z_{sat}$ are initialized to the values that exactly reproduce the SNM energy 
and pressure of FSU2 at $n=4 n_{sat}$ [see Section \ref{sec:lab} and Eq.(\ref{matrix_inv2})].
The two extremities of density points chosen to obtain the ``$\beta$-eq." and ``$\beta$-eq.+lab" 
solutions are the central densities corresponding to NSs with mass $1.4 M_{\odot}$ and $2.01 M_{\odot}$ in 
FSU2, with two more equispaced points in between for the former {\it i.e.} ``$\beta$-eq.". 
The coherence between the three lines in the upper panel of Fig. \ref{post1}
shows the absolute compatibility between FSU2 and its meta-model equivalents, as far as the 
$\beta$-equilibrium EoS is concerned.
However, if we turn to the composition (Fig. \ref{post1}(b)) , we observe a strong deviation 
between the model (black line) compared to the ``$\beta$-eq." solution (red dashed line), 
pointing towards multiple solutions of the   $\beta$-equilibrium equation.}

{The corresponding Bayesian analysis ``post-1" also supports
this toy model calculation. In Figure \ref{post1},  the  68\% and 95\%   posteriors
on the pressure and proton fraction, along with the  99\% prior are plotted
as  green, blue and orange bands, respectively.
We observe that the posterior for pressure is correctly centered on the reference model EoS, 
but the uncertainty in the composition covers the whole physical range of
$x_p$, with the highest probability concentrated on the solution given by the functional solution  
that is not compatible with the reference FSU2 functional. The compatibility of the Bayesian 
analysis with the ``$\beta$-eq." solution is most likely due to the fact that we have 
arbitrarily centered around zero in the prior distribution of the unknown high order NMPs. 
However, a more educated guess is difficult to justify given the absence of direct constraints on 
those quantities.}
 
 \subsection{Constraining the $M$-$R$ relation}\label{sec:MR}
As already discussed in Section \ref{sec:Bayes}, realistic experimental constraints on the 
nuclear functional are given by the observation of integrated quantities such as the mass, 
the radius, or the tidal polarizability of NSs. We therefore turn to the post-2 protocol, 
where  three hypothetical radii measurements at fixed NS  
masses are imposed, centered on the FSU2 theoretical value: $R_{1.4M_{\odot}}=14.18\pm 
0.71$ km, $R_{1.7M_{\odot}}=13.79\pm 0.69$ km and  $R_{2.01M_{\odot}}=12.93\pm 0.65$ km. 
In Fig. \ref{post2}(a), these are denoted by the vertical black error bands on the FSU2 $M$-$R$
curve (black line). The corresponding 68\% and 95\%   posteriors
on the radii, {accompanied by 99\% prior are plotted
as  green, blue and orange bands, respectively.
{The accuracy on the posteriors obtained in the Bayesian analysis conditioned on $M$-$R$, 
is similar to the one obtained by directly imposing  the EoS behavior as in post-1.
\begin{figure}[h]{}
\includegraphics[height=1.7in,width=3.in]{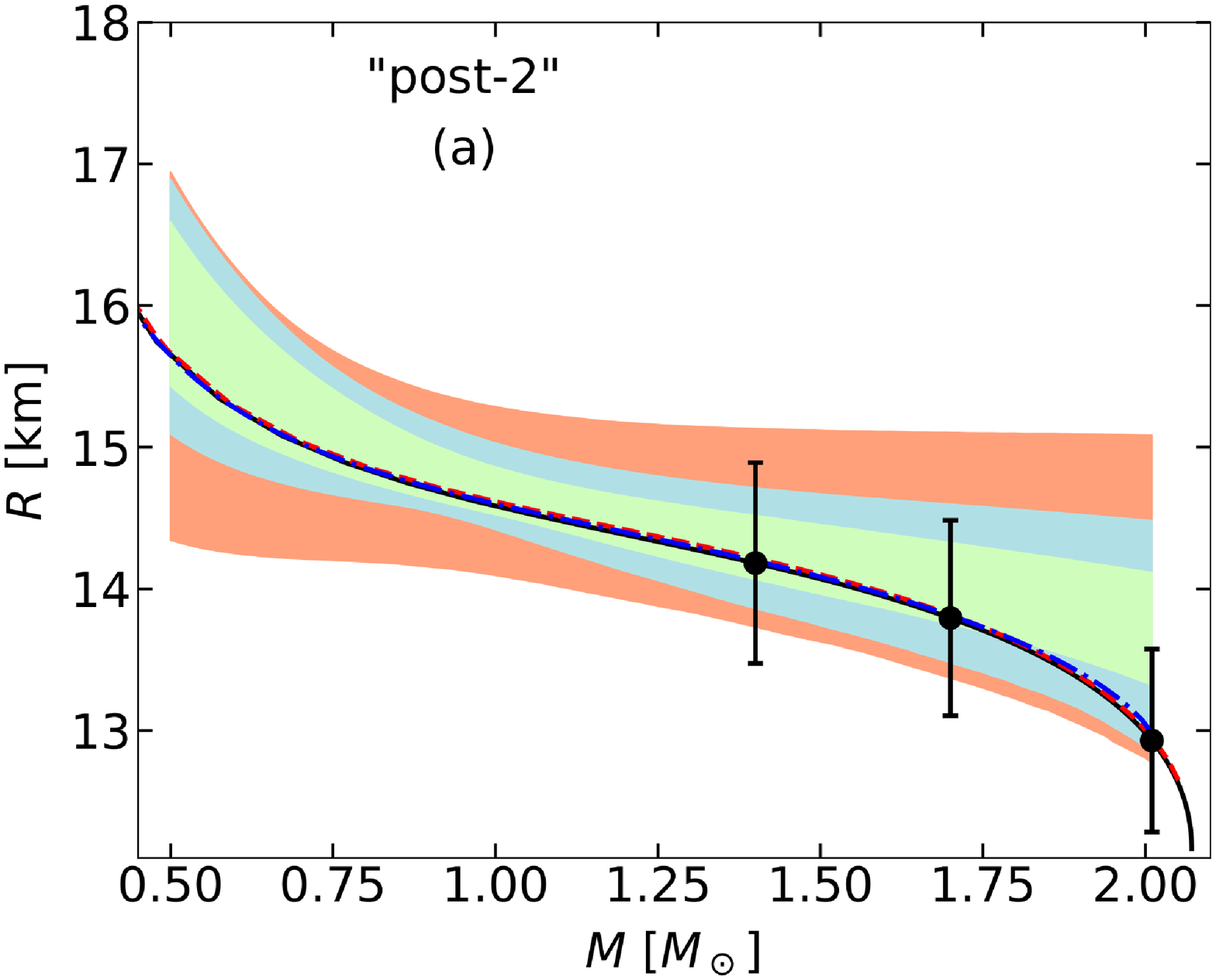}
\includegraphics[height=1.7in,width=3.in]{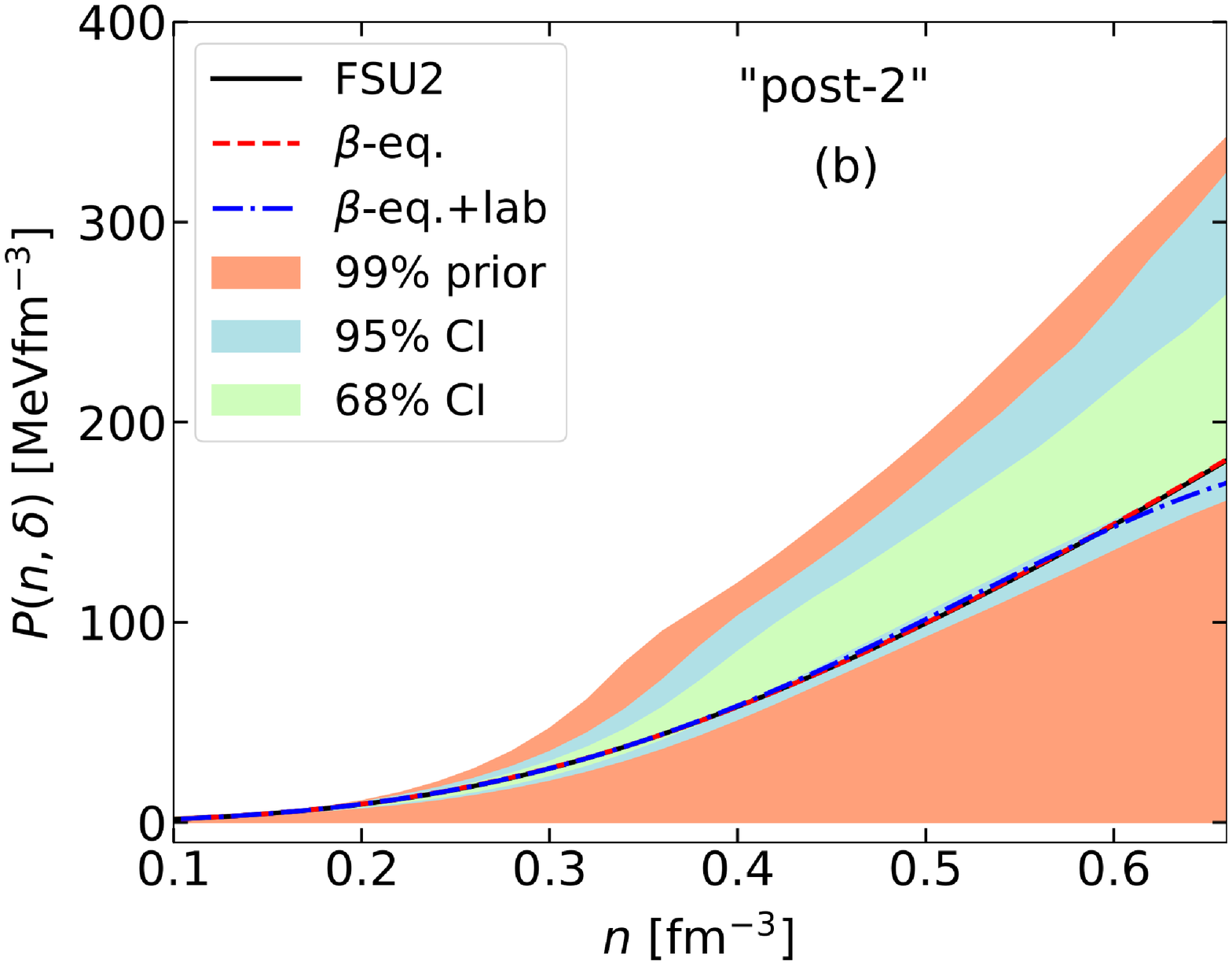}
\includegraphics[height=1.7in,width=3.in]{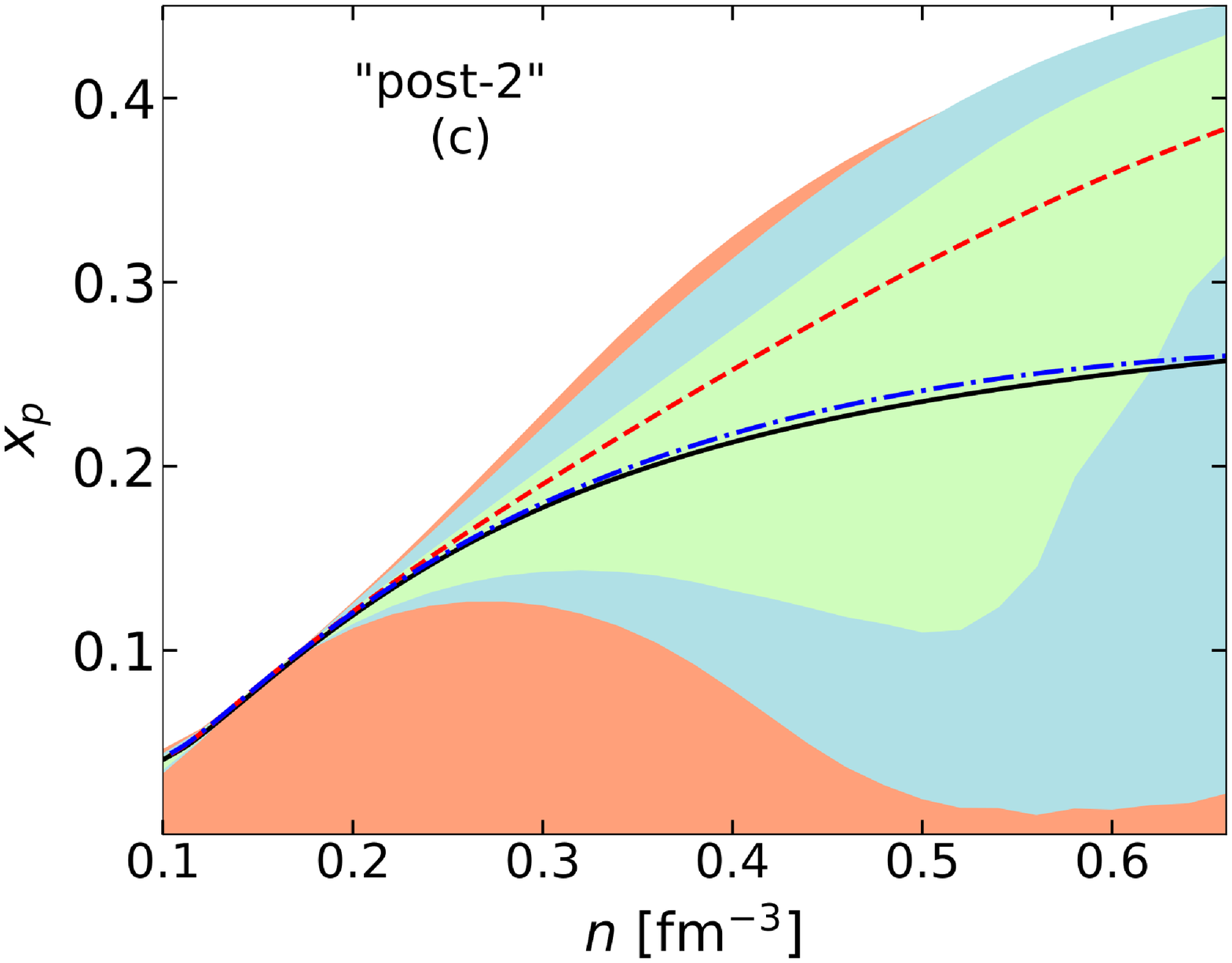}
\caption{\label{post2}
	(Color online) Radius versus mass (panel a), Pressure (panel b) and proton fraction (panel c) 
	versus  density for FSU2 and its meta-model equivalents,  along with its {prior and} posterior distributions
        with different confidence intervals (CI) calculated from ``post-2"  (see text for details).
}
\end{figure}
The 5\% error bar on the pseudo-observations converts into $\approx 20$\% 
uncertainty in the high density EoS, as displayed in Fig. \ref{post2}(b). This is due to 
the well known non-linearity of the relation between $M$-$R$ and EoS, as given by the 
TOV equation. We can also observe a systematic shift compared to the ``true'' values  
towards higher pressure as the density increases, with a corresponding shift in the radius,
increasing with increasing NS mass. This behavior is expected in any Bayesian analysis, 
if the prior is not centered on the true value.} 
{We can also see that} the width of the posterior $M$-$R$ band is thinner than the hypothetical measurement.
{This can be appreciated from the fact that the separate radii measurements effectively 
constrain the same parameters, leading to a stronger constraint than any individual 
measurement. Moreover,}
the strong assumption of an exact knowledge of the low density physics encoded in the
lower order parameters in Eq. (\ref{meta}) strongly constrains the EoS up to $\sim 2n_{sat}$,
as seen in Fig. \ref{post1}.  {The fact that only the higher order parameters are 
varied is not sufficient to eliminate the global bias observed, but it
contributes to narrow the posterior prediction.}}

{{All in all, the true EoS is recovered at the 1-$\sigma$ level, which can be considered 
as a very satisfactory EoS  determination. This is an illustration of the well know fact that 
a precise measurement of NS radii is an extremely powerful EoS estimator.} However, 
the posterior distribution of the proton fraction is very similar to the one observed in Figure \ref{post1}, and the uncertainty in the composition covers the whole physical range of $x_p$. This confirms the observations of Section \ref{sec:prho}, namely the uncertainty in the composition appears not to be due to the imprecision in the knowledge of the EoS, but rather to the degeneracy between very different energy functionals
that happen to lead to the same $\beta$-equilibrium trajectory due to cancellations between the symmetric matter $U_0$ and the symmetry $U_{sym}$ functionals
(see Eq. (\ref{meta})).  
}

 \subsection{Constraining the SNM and symmetry energy}\label{sec:lab}
The results of Fig. \ref{post2} suggest that, to eliminate the degeneracy and pin down the 
behavior of the reference model, independent additional constraints are needed from either 
the symmetric or the asymmetric part of the functional to complement the astrophysical observations. 
This argument is confirmed by the fact that the analytical EoS inversion presented in Section 
\ref{sec:inverting} systematically leads to the correct reproduction of the reference functional even 
in the composition, if the $\beta$-equilibrium information is only used to extract the NMPs 
associated to the symmetry energy. In particular, we consider a high density point that we fix 
for illustrative purposes at $n=4n_{sat}$, and impose the meta-functional in Eq. (\ref{meta}) 
to exactly reproduce the SNM energy and pressure of the reference model. This fixes the  
parameters $Q_{sat}$ and $Z_{sat}$ \cite{Margueron18a}. We then follow the same method of matrix 
inversion as in Eq.(\ref{matrix}) to extract $Q_{sym}$ and $Z_{sym}$ as the following,
\begin{equation}
        \frac{1}{6}x^3{\delta_\beta}^2Q_{sym}+\frac{1}{24}x^4{\delta_\beta}^2Z_{sym}
        =e_{th}(n,\delta_\beta)-e_{meta}^{0,sat}(n,{\delta_\beta}),
\label{matrix_inv2}
\end{equation}
where $e_{meta}^{0,sat}$ is the functional $e_{meta}$ from Eq. (\ref{meta}) obtained by using the
lower order NMPs from the input functional, $Q_{sat}$ and $Z_{sat}$ from the method described
above and  $Q_{sym}=Z_{sym}=0$.
Here, we need only two density points, which we have chosen as the central densities of
FSU2 corresponding to NSs of masses 1.4$M_{\odot}$ and 2.01$M_{\odot}$.
As expected, among the different possible solutions of the $\beta$-equilibrium equation, 
this strategy selects the meta-functional that correctly reproduces the reference model 
both in the isoscalar and in the isovector sector (``$\beta$-eq.+lab"), represented by 
dot-dashed blue lines in Figs.\ref{post1}-\ref{post2}.

\begin{figure}[t]{}
\includegraphics[height=3.0in,width=2.8in,angle=-90]{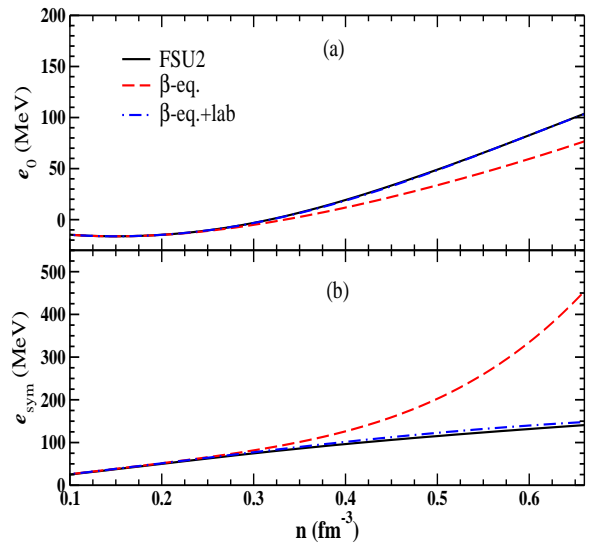}
\caption{\label{fsu2snmsym}
	(Color online) Energy for symmetric nuclear matter $e_0$ (panel a) and symmetry energy 
	$e_{sym}$ (panel b) as a function of number density for FSU2 and its meta-model 
	equivalents ``$\beta$-eq." and ``$\beta$-eq.+lab" are displayed. 
}
\end{figure}
In Fig. \ref{fsu2snmsym}, we display the behavior of SNM ($e_0$ in panel a) and 
symmetry energy ($e_{sym}$ in panel b) as a function of number density $n$ for 
the reference model FSU2 in conjunction with ``$\beta$-eq." and ``$\beta$-eq.+lab" 
meta-model equivalents represented by black solid, red-dashed and blue dash-dotted 
lines, respectively. One can notice that even though ``$\beta$-eq." solution 
doesn't match with $e_0$ and $e_{sym}$ of FSU2, it perfectly reproduces the 
$\beta$-equilibrium pressure (Fig. \ref{post1}(a)). This explains the mismatch 
between the red-dashed and black solid lines in the composition (Fig. \ref{post1}(b) 
and Fig. \ref{post2}(c)). However, for obvious reasons ``$\beta$-eq.+lab" 
solution agrees everywhere with FSU2 in Figs.\ref{post1}-\ref{post2} as it correctly 
reproduces the SNM and symmetry energy behavior separately (see Fig. \ref{fsu2snmsym}).

Information on the energy and pressure  of  symmetric
matter  at suprasaturation density is expected from upcoming measurements in 
relativistic heavy-ion collisions \cite{Adamczewski-Musch20}.
However, we expect those constraints to be affected by considerable uncertainties like 
any other measurements. To understand their effect on the determination of the composition,  we perform a similar
Bayesian analysis as before (``post-3"), by complementing the pseudo-observations of
radii introduced earlier, with a hypothetical SNM pressure observation at $4n_{sat}$
calculated from FSU2 with 10\% precision {\it i.e.} $P_{SNM}(0.6{\rm \ fm}^{-3})=125.14
\pm 12.51$ MeV.

\begin{figure}[]{}
\includegraphics[height=3.3in,width=3.in]{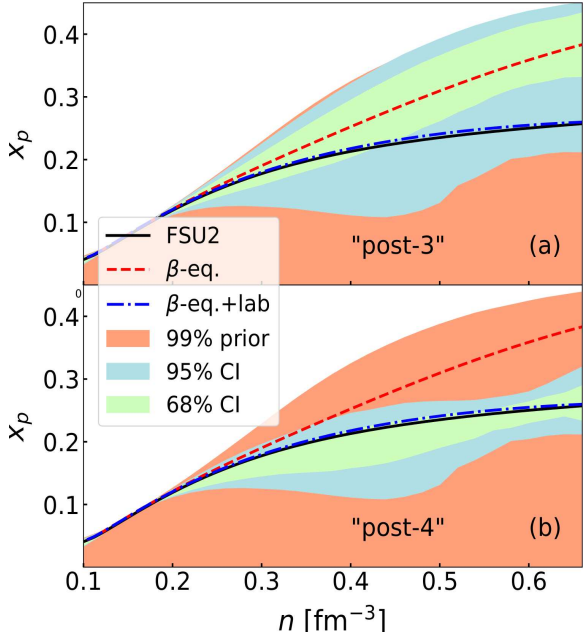}
\caption{\label{xpall}
(Color online) Proton fraction $x_p$ as a function of density to compare the distributions
        ``post-3" (panel (a)) and ``post-4"  (panel (b)) (see text for details).
}
\end{figure}
The resulting posterior  proton fraction $x_p$ as a function of density 
$n$ is displayed in Fig. \ref{xpall}(a).  One can notice that the
 distribution is less disperse, but it is still centered around the wrong composition
 described earlier (dashed red line).
The mismatch  can be understood as follows.
The energy of the $\beta$-equilibrated matter
in Eq. (\ref{eq:beta}) is determined by an
interplay between energy of SNM and symmetry energy. 
{Those quantities are not independent, since the relative proportion of the two 
is imposed by the $\beta$-equilibrium equation, but the latter possesses multiple solutions.}
   In the  case of the blue dash-dotted lines in Fig. \ref{xpall}, the symmetric matter NMPs  
are fitted exactly to their true values, which forces the symmetry energy and $x_p$
to take their model values (see Fig. \ref{fsu2snmsym}). 
But in ``post-3", the SNM is not fixed exactly,
which leaves a leeway for errors to creep in and get transferred further to
symmetry energy and $x_p$ because of the intrinsic degeneracy between the two parts of the functional. 
{This interpretation is in agreement with previous studies \cite{Xie19,Xie20,Li21}.}
To {better quantify the statement}, we add another constraint on top of ``post-3" by further imposing
the symmetry energy of FSU2 at $4n_{sat}$ as: $e_{sym}(0.6{\rm \ fm}^{-3})= 131.65\pm13.{16}$
MeV, which is coined as ``post-4". We plot the proton
fraction $x_p$ as a function of density for ``post-4" in Fig. \ref{xpall}(b),
and the 68\% posterior of the Bayesian analysis gets aligned with the true
composition, {\it albeit} with a non-negligible amount of uncertainty associated with it.

The different sensitivity of $x_p$ to $e_{\beta}$, $e_0$ and
$e_{sym}$ can be {qualitatively} understood via {analytical uncertainty} propagation.
If we employ a simple parabolic approximation for the symmetry energy
$e_{\beta}\approx e_0+ e_{sym}\delta_\beta^2 $,
 Eq. (\ref{eq:beta}) can be written in terms of the proton fraction as,
\begin{eqnarray}
        4(1-2x_p)e_{sym}-\Delta m_{np}=\mu_e(x_p).
\label{betaeqm}
\end{eqnarray}
In the ultra-relativistic approximation for the electron gas,
one can easily write the uncertainty in proton fraction $\Delta x_p$
in terms of uncertainties on $\beta$-equilibrium energy $\Delta e_{\beta}$ and
SNM energy $\Delta e_0$, or on symmetry energy $\Delta e_{sym}$ as
\begin{equation}
\Delta x_p \approx \frac{12x_p(\Delta e_{\beta}+\Delta e_0)}{\mu_e|1-8x_p|}
        \approx \frac{12(1-2x_p)^2x_p\Delta e_{sym}}{\mu_e (4x_p+1)}.
\label{erreqn}
\end{equation}
\begin{figure}[t]{}
\includegraphics[height=3.0in,width=2.8in,angle=-90]{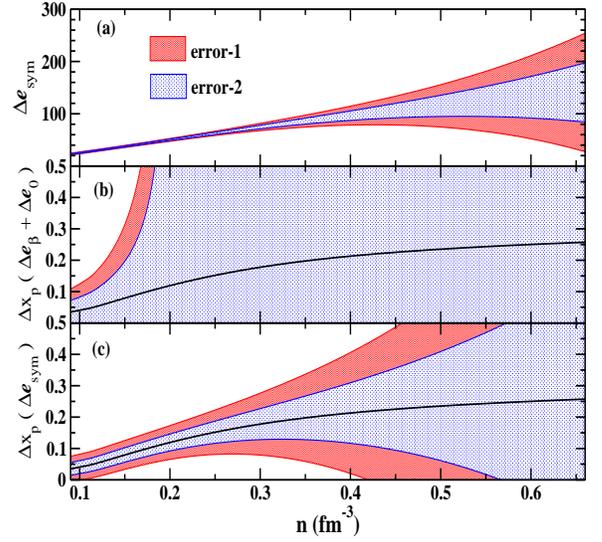}
\caption{\label{error}
(Color online) Uncertainty on the symmetry energy $\Delta e_{sym}$ as a function
        of density (panel (a)), its consequence on the uncertainties of proton
        fraction $\Delta x_p$ as a function of density (panel (c)) along with
        $\Delta x_p$ caused due to uncertainties in $\beta$-equilibrated energy
        $\Delta e_{\beta}$ and SNM energy $\Delta e_0$ (panel (b)),
        all obtained from Eq. (\ref{erreqn}) (see text for details).
}
\end{figure}
The first equality in Eq. (\ref{erreqn}) reflects the situation of ``post-3"
and the second one of ``post-4".
As a qualitative estimate of the present uncertainty in the symmetry energy,
we consider the uncertainties in the NMPs  proposed in Table IX of Ref. \cite{Margueron18a}.
In Fig. \ref{error}(a) we plot this (red band) as a function of density $n$.
The thinner blue band is simply half of it. Concerning $\Delta e_{\beta}$ and
$\Delta e_0$, we take the conservative choice $\Delta e_{\beta} + \Delta e_0=\Delta
e_{sym}$, such as to concentrate on the different uncertainty propagation due to
the form of the $\beta$-equilibrium equation, independent of the actual precision
of the measurements. The results are shown in Fig. \ref{error}(b) and (c).
One can see that the error band on
$\Delta x_p$ spans the whole space in Fig. \ref{error}(b), at variance with Fig. \ref{error}(c).
This can qualitatively explain the findings of our Bayesian analyses ``post-3"
and  ``post-4" of Fig. \ref{xpall},
pointing to the fact that the knowledge on symmetry energy at high
density plays stronger role in determining the proton fraction  inside the core
of a NS, than an equivalent information on SNM. {It is worthwhile to pose a
caution on the information derived from Eq. \ref{erreqn}. The
apparent divergence in $\Delta x_p$ at $x_p=0.125$ is an artifact of the simplistic 
way to extract the error from the {approximate
$\beta$-equilibrium relation Eq.(\ref{betaeqm})}. However, a smooth removal of the divergence still
engulfs the whole available space of proton fraction $x_p$ in Fig. \ref{error}(b).}

\section{Summary and Conclusions}\label{sec:concl}
In summary, we have analyzed the amount of empirical information which is needed
to decipher the composition of the core of a catalyzed non-rotating neutron star
within the nucleonic hypothesis. Using both a quasi-analytical inversion of
the $\beta$-equilibrium equation, and a more quantitative Bayesian analysis,
we have shown that one can land in a
wrong conclusion on the composition relying only on astrophysical informations
coming from static properties of neutron star.

In an ideal situation,
where one knows all the NMPs up to second order exactly, the composition can be extracted
by providing exact information on the energy per nucleon of SNM at a single high density point.
However, if uncertainties are taken into account even moderately, the core proton fraction
is fully unconstrained unless  the astrophysical observations are complemented
with accurate measurements on the symmetry energy at high density.

Finally, it is important to mention that this study gives only a lower limit on the
accessible uncertainty to the composition of neutron star interior. Further uncertainties
will arise from the possible incorporation of exotic degrees of freedom
{\it e.g.} hyperons or the deconfined quark matter, which are kept out in the present study.

{\it Acknowledgement-} The authors acknowledge partial support from
the IN2P3 Master Project “NewMAC”.


\end{document}